
\input harvmac.tex
\Title{CTP/TAMU-17/93}{Four-Dimensional String/String Duality
\footnote{$^\dagger$}
{Work supported in part by NSF grant PHY-9106593.}}

\centerline{M.~J.~Duff and
Ramzi R.~ Khuri\footnote{$^*$}{Supported by a World Laboratory Fellowship.}}
\bigskip\centerline{Center for Theoretical Physics}
\centerline{Texas A\&M University}\centerline{College Station, TX 77843}

\vskip .3in
We present supersymmetric soliton solutions of the four-dimensional heterotic
string corresponding to monopoles, strings and domain walls. These solutions
admit the $D=10$ interpretation of a fivebrane wrapped around $5$, $4$ or $3$
of the $6$ toroidally compactified dimensions and are arguably exact to all
orders in $\alpha'$. The solitonic string solution exhibits an $SL(2,Z)$
{\it strong/weak coupling} duality which however corresponds to an $SL(2,Z)$
{\it target space} duality of the fundamental string.

\Date{4/93}


\lref\monto {C. Montonen and D. Olive, Phys. Lett. {\bf B72} (1977) 117.}

\lref\sing {M. J. Duff, R. R. Khuri and J. X. Lu, Nucl. Phys. {\bf B377}
(1992) 281.}

\lref\rey{S.~J.~Rey, Phys. Rev. {\bf D43} (1991) 526.}

\lref\inst{R.~R.~Khuri, Phys. Lett. {\bf B259} (1991) 261.}

\lref\mono{R.~R.~Khuri, Phys. Lett. {\bf B294} (1992) 325.}

\lref\monex{R.~R.~Khuri, Nucl. Phys. {\bf B387} (1992) 315.}

\lref\monin{R.~R.~Khuri, Phys. Rev. {\bf D46} (1992) 4526.}

\lref\dghrr{A.~Dabholkar, G.~Gibbons, J.~A.~Harvey and F.~Ruiz Ruiz,
Nucl. Phys. {\bf B340} (1990) 33.}

\lref\fbrane{M.~J.~Duff and J.~X.~Lu, Nucl. Phys. {\bf B354} (1991) 141.}

\lref\duality{M.~J.~Duff and J.~X.~Lu, Nucl. Phys. {\bf B354} (1991) 129.}

\lref\dualtwo{M.~J.~Duff and J.~X.~Lu, Class. Quant. Gravity {\bf 9} (1992) 1.}

\lref\cqdual{M.~J.~Duff and J.~X.~Lu, Nucl. Phys. {\bf B357} (1991) 534.}

\lref\dfluhs{M.~J.~Duff and J.~X.~Lu, Phys. Rev. Lett. {\bf 66}
(1991) 1402.}

\lref\chsone{C.~G.~Callan, J.~A.~Harvey and A.~Strominger, Nucl. Phys.
{\bf B359} (1991) 611.}

\lref\chstwo{C.~G.~Callan, J.~A.~Harvey and A.~Strominger, Nucl. Phys.
{\bf B367} (1991) 60.}

\lref\strom{A.~Strominger, Nucl. Phys. {\bf B343} (1990) 167.}

\lref\duff{M.~J.~Duff, Class. Quant. Grav. {\bf 5} (1988).}

\lref\jjj{J.~P.~Gauntlett, J.~A.~Harvey and J.~T.~Liu, EFI-92-67,
IFP-434-UNC.}

\lref\chad{J.~M.~Charap and M.~J.~Duff, Phys. Lett. {\bf B69} (1977) 445.}

\lref\dixdp{J.~A.~Dixon, M.~J.~Duff and J.~C.~Plefka, Phys. Rev. Lett. {\bf 69}
(1992) 3009.}

\lref\senone{A.~Sen, TIFR-TH-92-41.}

\lref\sentwo{A.~Sen, TIFR-TH-93-03.}

\lref\schsen{J.~H.~Schwarz and A.~Sen, NSF-ITP-93-46, CALT-68-1863,
TIFR-TH-93-19.}

\lref\dhis{M.~J.~Duff, P.~Howe, T.~Inami and K.~S.~Stelle, Phys. Lett.
{\bf B191} (1987) 70.}

\lref\dipss{M.~J.~Duff, T.~Inami, C.~N.~Pope, E.~Sezgin and K.~S.~Stelle,
Nucl. Phys. {\bf B297} (1988) 515.}

\lref\fujku{K.~Fujikawa and J.~Kubo, Nucl. Phys. {\bf B356} (1991) 208.}

\lref\hsla{H.~La, CTP-TAMU-52/92.}

\lref\cvet{M.~Cveti\v c, UPR-560-T.}

\lref\sch{J.~Schwarz, CALT-68-1815.}

\lref\filq{A.~Font, L.~Ib\'a\~nez, D.~Lust and F.~Quevedo, Phys. Lett.
{\bf B249} (1990) 35.}

\lref\bin{P.~Bin\'etruy, NSF-ITP-93-60.}

\def\sqr#1#2{{\vbox{\hrule height.#2pt\hbox{\vrule width
.#2pt height#1pt \kern#1pt\vrule width.#2pt}\hrule height.#2pt}}}
\def\Box{\mathchoice\sqr64\sqr64\sqr{4.2}3\sqr33}
\def\cone{R^\mu{}_{\nu\rho\sigma}}
\def\gcone{\hat R^\mu{}_{\nu\rho\sigma}}

\def\met {g_{\mu\nu}}

\newsec{Introduction}

A major problem in string theory is to go beyond a weak-coupling perturbation
expansion. A possible approach to this problem is provided by the
string/fivebrane duality conjecture \refs{\duff,\strom}, which states that,
in their critical
spacetime dimension $D=10$, superstrings (extended objects with one spatial
dimension) are dual to superfivebranes (extended objects with five spatial
dimensions). There is now a good deal of evidence in favor of this idea, which
may be divided into: Poincare duality \duff, strong/weak coupling duality
\refs{\strom\duality\fbrane\chsone\chstwo\dfluhs{--}\dualtwo}, singularity
structure duality \sing\ and classical/quantum duality
\refs{\cqdual,\dixdp}. Most of these discussions have focused on the $D=10$
heterotic string and its dual counterpart the $D=10$ heterotic fivebrane,
but in this paper we wish to examine the four-dimensional consequences.

That the field theory limit of the $D=10$ heterotic string admits as a soliton
a heterotic fivebrane \duff\ was first pointed out by Strominger \strom. He
went on to suggest a strong/weak coupling duality between the string and the
fivebrane in analogy with the Montonen-Olive strong/weak coupling conjecture
in four-dimensional super Yang-Mills theories \monto. This strong/weak
coupling was subsequently confirmed from the point of view of Poincare
duality in \duality. There it was shown that just as the string loop
expansion parameter is given by
\eqn\stloop{g({\rm string})=e^{\phi_0},}
where $\phi_0$ is the dilaton VEV, so the analogous fivebrane parameter
is given by
\eqn\fbloop{g({\rm fivebrane})=e^{-\phi_0/3},}
and hence
\eqn\stfbdual{g({\rm fivebrane})=g({\rm string})^{-1/3}.}

In the same paper \strom, Strominger pointed out that after toroidal
compactification to four dimensions, the fivebrane would appear as either a
0-brane, a 1-brane or a 2-brane, depending in how the fivebrane wraps around
the compactified directions \refs{\dhis\dipss{--}\fujku}.
Thus it ought to be possible to find soliton solutions directly from the
four-dimensional string corresponding to monopoles (0-branes), strings
(1-branes) and domain walls (2-branes).
Such fivebrane-inspired supersymmetric monopoles have already been found
\refs{\mono\monex{--}\jjj}, and here we discuss the string and domain
wall.
(The possibility of monopole
solutions in this context was anticipated in \inst, where a monopole-like
solution in the massless fields sector of the bosonic string was discovered.)

To find these multi-string and multi-domain wall solutions we shall follow
the procedure outlined in \refs{\mono,\monex}\ for monopoles, where it was
argued that exact solutions of the heterotic string could be obtained by
modifying the 't Hooft ansatz for the Yang-Mills instanton. We shall present
them
from both the $D=10$ and $D=4$ points of view. As with the fivebrane and the
monopole, there are three types of string and domain wall solutions:
neutral, gauge and symmetric. Following the arguments of
\refs{\chsone,\chstwo}, all symmetric solutions correspond to $(4,4)$
supersymmetry on the worldsheet of the fundamental string and are thus
presumably exact to all orders in $\alpha'$.

Of particular interest is the solitonic string, since its couplings to the
background fields of supergravity are the same as those of the fundamental
string except that the dilaton/axion field $S$ is replaced by the modulus
field $T$. Thus under string/fivebrane duality, the $SL(2,Z)$ strong/weak
coupling duality trades places with the $SL(2,Z)$ target space duality,
in accordance with recent observations of Schwarz and Sen \schsen\ and
Bin\'etruy \bin.

\newsec{The General Ansatz}

We first summarize the 't Hooft ansatz for the Yang-Mills instanton.
Consider the four-dimensional Euclidean action
\eqn\eucym{ S=-{1\over 2g^2}\int d^4x {\rm Tr} F_{\mu\nu}F^{\mu\nu},
\qquad\qquad \mu,\nu =1,2,3,4.}
For gauge group $SU(2)$, the fields may be written as $A_\mu=(g/2i)
\sigma^a A_\mu^a$ and $F_{\mu\nu}=(g/2i)\sigma^a F_{\mu\nu}^a$\ \
(where $\sigma^a$, $a=1,2,3$ are the $2\times 2$ Pauli matrices).
A self-dual solution (but not the most general one) to the equation of motion
of this action is given by the 't Hooft ansatz
\eqn\hfan{A_\mu=i \overline{\Sigma}_{\mu\nu}\partial_\nu \ln f,}
where $\overline{\Sigma}_{\mu\nu}=\overline{\eta}^{i\mu\nu}(\sigma^i/2)$
for $i=1,2,3$, where
\eqn\hfantwo{\eqalign{\overline{\eta}^{i\mu\nu}=-\overline{\eta}^{i\nu\mu}
&=\epsilon^{i\mu\nu},\qquad\qquad \mu,\nu=1,2,3,\cr
&=-\delta^{i\mu},\qquad\qquad \nu=4  \cr}}
and where $f^{-1}\Box\ f=0$. The ansatz for the anti-self-dual solution
is similar, with the $\delta$-term in \hfantwo\ changing sign.
{}From this ansatz, depending on how many of the four coordinates $f$ is
allowed to depend and depending on whether we compactify, we shall obtain
$D=10$ multi-fivebrane and $D=4$ multi-monopole,
multi-string and multi-domain wall solutions. We will discuss these four
cases in the next section. In this section, we do not specify the precise
form of $f$ or the dilaton function, but show that the derivation of the
solution and most of the arguments used to demonstrate
the exactness of the heterotic solution are equally valid for any $f$
satisfying $f^{-1}\Box\ f=0$.

It turns out that there is an analog to the 't Hooft ansatz for the Yang-Mills
instanton in the gravitational sector of the string, namely the axionic
instanton \rey. In its simplest form, this instanton appears as a solution for
the massless fields of the bosonic string \inst. The identical instanton
structure arises in all supersymmetric multi-fivebrane solutions
\refs{\fbrane,\chsone}, in particular in the tree-level neutral solution
\fbrane:
\eqn\fbsol{\eqalign{\met&=e^{2\phi}\delta_{\mu\nu}\qquad \mu,\nu=1,2,3,4,\cr
g_{ab}&=\eta_{ab}\qquad\quad   a,b=0,5,...,9,\cr
H_{\mu\nu\lambda}&=\pm 2\epsilon_{\mu\nu\lambda\sigma}\partial^\sigma\phi
\qquad \mu,\nu,\lambda,\sigma=1,2,3,4,\cr}}
with $e^{-2\phi}\Box\ e^{2\phi}=0$. The D'Alembertian refers to the
four-dimensional subspace $\mu,\nu,\lambda,\sigma=1,2,3,4$ and $\phi$ is taken
to be independent of $(x^0,x^5,x^6,x^7,x^8,x^9)$. For zero background fermionic
fields the above solution breaks half the spacetime supersymmetries.

The generalized curvature of this solution was shown \refs{\inst,\monin}\
to possess (anti) self-dual structure similar to that of the 't Hooft ansatz.
To see this we define a generalized curvature $\gcone$ in terms of the standard
curvature $\cone$ and $H_{\mu\alpha\beta}$:
\eqn\gcurv{\gcone=\cone+{1\over 2}\left(\nabla_\sigma H^\mu{}_{\nu\rho}-
\nabla_\rho H^\mu{}_{\nu\sigma}\right)+
{1\over 4}\left(H^\lambda{}_{\nu\rho}H^\mu{}_{\sigma\lambda}
- H^\lambda{}_{\nu\sigma} H^\mu{}_{\rho\lambda}\right).}
One can also define $\gcone$ as the Riemann tensor generated
by the generalized Christoffel symbols $\hat\Gamma^\mu_{\alpha\beta}$,
where $\hat\Gamma^\mu_{\alpha\beta}=\Gamma^\mu_{\alpha\beta}
-(1/2) H^\mu{}_{\alpha\beta}$.
The crucial observation for obtaining higher-loop and even exact solutions
is the following. For any solution given by \fbsol,
we can express the generalized curvature in terms of the dilaton field as \inst
\eqn\gcurvtwo{\eqalign{\gcone&=
\delta_{\mu\sigma}\nabla_\rho\nabla_\nu\phi
-\delta_{\mu\rho}\nabla_\sigma\nabla_\nu\phi
+\delta_{\nu\rho}\nabla_\sigma\nabla_\mu\phi
-\delta_{\nu\sigma}\nabla_\rho\nabla_\mu\phi \cr
&\pm\epsilon_{\mu\nu\rho\lambda}\nabla_\sigma\nabla_\lambda\phi
\mp\epsilon_{\mu\nu\sigma\lambda}\nabla_\rho\nabla_\lambda\phi.\cr}}
It easily follows that
\eqn\axin{\gcone=\mp {1\over 2} \epsilon_{\rho\sigma}{}^{\lambda\gamma}
\hat R^\mu{}_{\nu\lambda\gamma}.}
So the (anti) self-duality appears in the gravitational sector of the string
in terms of its generalized curvature.

We now turn to the exact heterotic solution. The tree-level supersymmetric
vacuum equations for the heterotic string are given by
\eqn\sseq{\eqalign{\delta\psi_M&=\left(\nabla_M-{\textstyle {1\over 4}}H_{MAB}
\Gamma^{AB}\right)\epsilon=0,\cr
\delta\lambda&=\left(\Gamma^A\partial_A\phi-{\textstyle{1\over 6}}
H_{ABC}\Gamma^{ABC}\right)\epsilon=0,\cr
\delta\chi&=F_{AB}\Gamma^{AB}\epsilon=0, \cr}}
where $A,B,C,M=0,1,2,...,9$ and
where $\psi_M,\ \lambda$ and $\chi$ are the gravitino, dilatino and gaugino
fields. The Bianchi identity is given by
\eqn\bianchi{dH={\alpha'\over 4} \left({\rm tr} R\wedge R-{1\over 30}{\rm Tr}
F\wedge F\right).}
The $(9+1)$-dimensional Majorana-Weyl fermions decompose into
chiral spinors according to $SO(9,1)\supset SO(5,1) \otimes SO(4)$ for
the $M^{9,1}\to M^{5,1}\times M^4$ decomposition. Then \fbsol\ with
arbitrary dilaton and with constant chiral spinors $\epsilon_\pm$ solves the
supersymmetry equations with zero background fermi fields provided the YM gauge
field satisfies the instanton (anti) self-duality condition \strom
\eqn\ymin{F_{\mu\nu}=\pm {1\over 2}\epsilon_{\mu\nu}{}^{\lambda\sigma}
F_{\lambda\sigma}.}
In the absence of a gauge sector, the multi-fivebrane solution is identical to
the ``neutral" tree-level solution shown in \fbsol. A perturbative ``gauge"
fivebrane solution was found in \strom.
An exact solution is obtained as follows. Define a generalized connection by
\eqn\gcon{\Omega^{AB}_{\pm M}=\omega^{AB}_M\pm H^{AB}_M }
in an SU(2) subgroup of the gauge group, and equate it to the gauge
connection $A_\mu$ \chad\ so that the corresponding curvature $R(\Omega_{\pm})$
cancels against the Yang-Mills field strength $F$ and $dH=0$.
For $e^{-2\phi}\Box\ e^{2\phi}=0$ (or $e^{2\phi} = e^{2\phi_0} f$) the
curvature of the generalized connection can be written in
terms of the dilaton as in \gcurvtwo\ from which it follows that both $F$ and
$R$ are (anti) self-dual. This solution becomes exact since
$A_\mu=\Omega_{\pm\mu}$ implies that all the higher order corrections vanish
\chsone. The self-dual solution for the gauge connection is then
given by the 't Hooft ansatz. So the heterotic solution combines a YM instanton
in the gauge sector with an axionic instanton in the gravity sector. In
addition, the heterotic solution has finite action. Further arguments
supporting the exactness of this solution based on $(4,4)$ worldsheet
supersymmetry are shown in \chsone. Note that at no point in this discussion
do we refer to the specific form of $f$, so that all of the above
arguments apply for an arbitrary solution of $f^{-1}\Box\ f=0$.

\newsec{Monopoles, Strings and Domain Walls}

We now go back to the 't Hooft ansatz \eucym-\hfantwo\ and solve the equation
$f^{-1}\Box\ f=0$. If we take $f$ to depend on all four coordinates we obtain
a multi-instanton solution
\eqn\fins{ f_I=1+\sum_{i=1}^N{\rho_i^2\over |\vec x - \vec a_i|^2},}
where $\rho_i^2$ is the instanton scale size and $\vec a_i$ the location in
four-space of the $i$th instanton. For $e^{2\phi} = e^{2\phi_0} f_I$,
and assuming no dimensions are compactified, we obtain
from \fbsol\ the neutral fivebrane of \fbrane\ and the exact heterotic
fivebrane
of \refs{\chsone,\chstwo}\ in $D=10$. The solitonic fivebrane tension
$\widetilde{T_6}$ is related to the fundamental string tension $T_2$
($=1/2\pi\alpha'$) by the Dirac quantization condition \duality\
\eqn\stfbdual{\kappa_{10}^2 \widetilde{T_6} T_2 =n\pi,}
where $n$ is an integer and where $\kappa_{10}^2$ is the $D=10$
gravitational constant. This implies $\rho_i^2=e^{-2\phi_0}n_i\alpha'$,
where $n_i$ are integers. Near each source the solution is described by
an exact conformal field theory \refs{\rey,\inst,\chsone}.

Instead, let us single out a direction in the transverse four-space (say $x^4$)
and assume all fields are independent of this coordinate. Since all fields
are already independent of $x^5,x^6,x^7,x^8,x^9$, we may consistently assume
the $x^4,x^5,x^6,x^7,x^8,x^9$ are compactified on a six-dimensional torus,
where we shall take the $x^4$ circle to have circumference $Le^{-\phi_0}$
and the rest to have circumference $L$,
so that $\kappa_4^2=\kappa_{10}^2e^{\phi_0}/L^6$. Then the solution
for $f$ satisfying $f^{-1} \Box\ f=0$ has multi-monopole structure
\eqn\fmono{f_M=1+\sum_{i=1}^N{m_i\over |\vec x - \vec a_i|},}
where $m_i$ is proportional to the charge and $\vec a_i$ the location in the
three-space $(123)$
of the $i$th monopole. If we make the identification $\Phi\equiv A_4$ then the
lagrangian density may be rewritten as
\eqn\relag{F_{\mu\nu}^a F_{\mu\nu}^a =F_{jk}^a F_{jk}^a + 2F_{k4}^a F_{k4}^a
=F_{jk}^a F_{jk}^a + 2D_k \Phi^a D_k \Phi^a,}
where $j,k=1,2,3$.
We now go to $3+1$ space $(0123)$ with the Lagrangian density
\eqn\lagden{{\cal L}=-{1\over 4}G_{\alpha\beta}^a G^{\alpha\beta a} -{1\over 2}
D_\alpha \Phi^a D^\alpha \Phi^a,}
where $\alpha,\beta=0,1,2,3$.
It follows that the above multi-monopole ansatz is a static solution with
$A_0^a=0$ and all time derivatives vanish. The solution in $3+1$ dimensions
has the form
\eqn\monsol{\eqalign{\Phi^a&=\mp{1\over g}\delta^{aj}\partial_j \omega,\cr
A_k^a&={1\over g}\epsilon^{akj}\partial_j \omega,\cr}}
where $j,k=1,2,3$ and
where $\omega\equiv \ln f$ and $g$ is the YM coupling constant. This solution
represents a multi-monopole
configuration with sources at $\vec a_i, i=1,2...N$ \refs{\mono,\monex}. For
$e^{2\phi} = e^{2\phi_0} f_M$, we obtain from \fbsol\ a neutral monopole
solution and the exact heterotic monopole solution of \refs{\mono,\monex}.
The monopole strength is given by $\tilde g=\sqrt{2}\kappa_4 \widetilde{T_1}$,
where $\widetilde{T_1}=\widetilde{T_6} L^5$ obeys, from \stfbdual, the
quantization condition
\eqn\stmondual{e^{-\phi_0}\kappa_4^2 T_2 \widetilde{T_1}={n\pi\over L}.}
This implies $m_i=e^{-\phi_0}n_i\pi\alpha'/L$. Similarly the ``electric''
charge of the fundamental string is $e=\sqrt{2} \kappa_4 T_1$, where
$T_1=T_2 Le^{-\phi_0}$, and hence
\eqn\dirac{e\tilde g=2\pi n}
as expected. Unlike for the instanton, in the monopole case we
cannot identify the explicit coset conformal field theory near each source.
A noteworthy feature of this solution is that the divergences
from both gauge and gravitational sectors cancel to yield a finite lagrangian,
and finite soliton mass.

It is straightforward to reduce the multi-monopole solution to an explicit
solution in the four-dimensional space $(0123)$. The gauge field
reduction is exactly as above, i.e. we replace $A_4$ with the scalar
field $\Phi$. In the gravitational sector, the reduction from ten to
five dimensions is trivial, as the metric is flat in the subspace
$(56789)$. In going from five to four dimensions, one follows the usual
Kaluza-Klein procedure of replacing $g_{44}$ with a scalar field
$e^{-2\sigma}$. The tree-level effective action reduces in four dimensions to
\eqn\redmon{S_4={1\over 2\kappa_4^2}\int d^4 x \sqrt{-g} e^{-2\phi - \sigma}
\left( R + 4(\partial\phi)^2 + 4\partial\sigma\cdot\partial\phi -
e^{2\sigma} {M_{\alpha\beta}M^{\alpha\beta}\over 4} \right),}
where $\alpha,\beta=0,1,2,3$,
where $M_{\alpha\beta}=H_{\alpha\beta}=\partial_\alpha B_\beta -
\partial_\beta B_\alpha$, and where
$B_\alpha=B_{\alpha 4}$. The four-dimensional monopole solution for this
reduced
action is then given by
\eqn\redmsol{\eqalign{e^{2\phi}=e^{-2\sigma}&=e^{2\phi_0}\left(
1+\sum_{i=1}^N{m_i\over |\vec x - \vec a_i|}\right),\cr
ds^2&=-dt^2 + e^{2\phi}\left(dx_1^2 + dx_2^2 + dx_3^2\right),\cr
M_{ij}&=\pm\epsilon_{ijk}\partial_k e^{2\phi},\qquad i,j,k=1,2,3.\cr}}
Since the tree-level solution is exact, we need not reduce the
higher order corrections to the action.

We now modify the solution of the 't Hooft ansatz even further and choose
two directions in the four-space $(1234)$ (say $x^3$ and $x^4$) and assume all
fields are independent of both of these coordinates. We may now consistently
assume that $x^3,x^4,x^6,x^7,x^8,x^9$ are compactified on a six-dimensional
torus, where we shall take the $x^3$ and $x^4$ circles to have circumference
$Le^{-\phi_0}$ and the remainder to have circumference $L$, so that
$\kappa_4^2=\kappa_{10}^2 e^{2\phi_0}/L^6$. Then the solution
for $f$ satisfying $f^{-1} \Box\ f=0$ has multi-string structure
\eqn\fstring{f_S=1-\sum_{i=1}^N \lambda_i \ln |\vec x - \vec a_i|,}
where $\lambda_i$ is the charge per unit length and $\vec a_i$ the location in
the two-space $(12)$ of the $i$th string. If we make the identification
$\Phi\equiv A_4$ and $\Psi\equiv A_3$
then the lagrangian density for the above ansatz can be rewritten as
\eqn\slag{F_{\mu\nu}^a F_{\mu\nu}^a = F_{jk}^a F_{jk}^a +
2D_k \Phi^a D_k \Phi^a + 2D_k \Psi^a D_k \Psi^a ,}
where $j,k=1,2$. We now go to the $3+1$ space $(0125)$ with the lagrangian
density
\eqn\lagden{{\cal L}=-{1\over 4}G_{\rho\sigma}^a G^{\rho\sigma a} -{1\over 2}
D_\rho \Phi^a D^\rho \Phi^a -{1\over 2} D_\rho \Psi^a D^\rho \Psi^a,}
where $\rho,\sigma=0,1,2,5$.
It follows that the multi-string ansatz is a static solution with
$A_0^a=0$ and all time derivatives vanish. The solution in $3+1$ dimensions
has the form
\eqn\stsol{\eqalign{\Phi^a&=\mp{1\over g}\delta^{aj}\partial_j \omega,\cr
\Psi^k&={1\over g}\epsilon^{kj}\partial_j \omega,\cr
A_k^a&=-\delta^{a3}{1\over g}\epsilon^{kj}\partial_j \omega,\cr}}
where $j,k=1,2$ and where $\omega\equiv \ln f$. This solution represents a
multi-string configuration with sources at $\vec a_i, i=1,2...N$. By setting
$e^{2\phi} = e^{2\phi_0}f_S$, we obtain from \fbsol\ a neutral multi-string
solution and an exact heterotic multi-string solution. The solitonic string
tension $\widetilde{T_2}$ is given by $\widetilde{T_6} L^4$
and from \stfbdual\ is related to the fundamental string tension $T_2$ by
\eqn\ststdual{e^{-2\phi_0}\kappa_4^2 T_2 \widetilde{T_2}={n\pi\over L^2}.}
This implies $\lambda_i=n_i 2\pi\alpha'/L^2$.
Like the monopole and unlike the instanton, we cannot identify an explicit
coset conformal field theory near each source. Also like the monopole, the
lagrangian per unit length for the string solution is finite as a result of the
cancellation of divergences between the gauge and gravitational sectors.

As in the multi-monopole case, it is straightforward to reduce the multi-string
solution to a solution in the four-dimensional space $(0125)$. The gauge field
reduction is done in \stsol. In the gravitational sector, the reduction from
ten to six dimensions is trivial, as the metric is flat in the subspace
$(6789)$. In going from six to four dimensions, we compactify the $x_3$ and
$x_4$
directions and again follow the Kaluza-Klein procedure by replacing $g_{33}$
and $g_{44}$ with a scalar field $e^{-2\sigma}$. The tree-level effective
action reduces in four dimensions to
\eqn\redst{S_4={1\over 2\kappa_4^2}\int d^4 x \sqrt{-g} e^{-2\phi - 2\sigma}
\left( R + 4(\partial\phi)^2 + 8\partial\sigma\cdot\partial\phi +
2(\partial\sigma)^2 - e^{4\sigma} {N_\rho N^\rho\over 2} \right),}
where $\rho=0,1,2,5$, where $N_\rho=H_{\rho 34}=\partial_\rho B$, and where
$B=B_{34}$. The four-dimensional
string soliton solution for this reduced action is then given by
\eqn\redssol{\eqalign{e^{2\phi}=e^{-2\sigma}&=e^{2\phi_0}\left(
1-\sum_{i=1}^N \lambda_i \ln |\vec x - \vec a_i|\right),\cr
ds^2&=-dt^2 + dx_5^2 + e^{2\phi}\left(dx_1^2 + dx_2^2\right),\cr
N_i&=\pm \epsilon_{ij} \partial_j e^{2\phi}.\cr}}
Again since the tree-level solution is exact, we do not bother to reduce the
higher order corrections to the action.

We complete the family of solitons that can be obtained from the solutions
of the 't Hooft ansatz by demanding that $f$ depend on only one coordinate,
say $x^1$. We may now consistently assume that $x^2,x^3,x^4,x^7,x^8,x^9$ are
compactified on a six-dimensional torus, where we shall take the $x^2$,
$x^3$ and $x^4$ circles to have circumference $Le^{-\phi_0}$ and the rest to
have circumference $\kappa_4^2=\kappa_{10}^2e^{3\phi_0}/L^6$. Then the solution
of $f^{-1} \Box\ f=0$ has domain wall structure with the ``confining potential"
\eqn\fdom{f_D=1+\sum_{i=1}^N \Lambda_i |x_1-a_i|,}
where $\Lambda_i$ are constants. By setting
$e^{2\phi} = e^{2\phi_0} f_D$, we obtain from
\fbsol\ a neutral domain wall solution and an exact heterotic domain wall
solution. The solitonic domain wall
tension $\widetilde{T_3}$ is given by $\widetilde{T_6} L^3$
and from \stfbdual\ is related to the fundamental string tension $T_2$ by
\eqn\stdwdual{e^{-3\phi_0}\kappa_4^2 T_2 \widetilde{T_3}={n\pi\over L^3}.}
This implies $\Lambda_i=e^{\phi_0}n_i (2\pi)^2\alpha'/L^3$. Like the monopole
and string we cannot identify an explicit coset conformal field theory near
each source. Again the reduction to $D=4$ is straightforward. In the gauge
sector, the action reduces to YM + three scalar fields $\Phi$, $\Psi$ and
$\Pi$. For the spacetime $(0156)$ the solution for the fields is given by
\eqn\domsol{\eqalign{\Phi^1&=\mp {\Lambda\over g(1+\Lambda |x_1|)},\cr
\Psi^3&={\Lambda \over g(1+\Lambda |x_1|)},\cr
\Pi^2&=-{\Lambda \over g(1+\Lambda |x_1|)},\cr
A_\mu &=0,\cr}}
where $\mu=0,1,5,6$. In the gravitational sector the tree-level effective
action in $D=4$ has the form
\eqn\reddom{S_4={1\over 2\kappa^2}\int d^4 x \sqrt{-g} e^{-2\phi - 3\sigma}
\left( R + 4(\partial\phi)^2 + 12\partial\sigma\cdot\partial\phi +
6(\partial\sigma)^2 - e^{6\sigma} {P^2\over 2} \right),}
where $P=H_{234}$. The four-dimensional domain wall
solution for this reduced action is then given by
\eqn\reddomsol{\eqalign{e^{2\phi}=e^{-2\sigma}&=e^{2\phi_0}\left(1+\Lambda
|x_1|\right),\cr
ds^2&=-dt^2 + dx_5^2 + dx_6^2 + e^{2\phi} dx_1^2,\cr
P&=\Lambda \left(\Theta(x_1)-\Theta(-x_1)\right).\cr}}
Again since the tree-level solution is exact, we do not bother to reduce the
higher order corrections to the action. A trivial change of coordinates
reveals that the spacetime is, in fact, flat. Dilaton domain walls with a flat
spacetime have recently been discussed in a somewhat different context
in \refs{\hsla,\cvet}.

As for the fivebrane in $D=10$, the mass of the monopole, the mass per unit
length of the string and the mass per unit area of the domain wall saturate
a Bogomol'nyi bound with the topological charge. (In the case of the string
and domain, wall, however, we must follow \dghrr\ and extrapolate the
meaning of the ADM mass to non-asymptotically flat spacetimes.)

\newsec{String/String Duality}

Let us focus on the solitonic string configuration \redssol\ in the case of
a single source. In terms of the complex field
\eqn\tfield{\eqalign{T&=T_1+iT_2\cr &=B_{34}+ie^{-2\sigma} \cr
&=B_{34}+i\sqrt{{\rm det} g^S_{mn}}
\qquad m,n=3,4,6,7,8,9,\cr}}
where $g^S_{MN}$ is the string $\sigma$-model metric, the solution takes the
form (with $z=x_1+x_2$)
\eqn\tsol{\eqalign{T&={1\over 2\pi i}\ln {z\over r_0},\cr
ds^2&=-dt^2 + dx_5^2 - {1\over 2\pi} \ln{r\over r_0} dz d\overline z,\cr}}
whereas both the four-dimensional (shifted) dilaton $\eta=\phi + \sigma$
and the four-dimensional two-form $B_{\mu\nu}$ are zero. In terms of the
canonical metric $g_{\mu\nu}$, $T_1$ and $T_2$, the relevant part of the
action takes the form
\eqn\sgtt{S_4={1\over 2\kappa_4^2}\int d^4 x\sqrt{-g}
\left( R - {1\over 2T_2^2}g^{\mu\nu}\partial_\mu T \partial_\nu \overline T
\right)}
and is invariant under the $SL(2,R)$ transformation
\eqn\sltwor{T\to {aT+b\over cT+d},\qquad ad-bc=1.}
The discrete subgroup $SL(2,Z)$, for which $a$, $b$, $c$ and $d$ are
integers, is just a subgroup of the $O(6,22;Z)$ {\it target space duality},
which can be shown to be an exact symmetry of the compactified string theory
at each order of the string loop perturbation expansion.

This $SL(2,Z)$ is to be contrasted with the $SL(2,Z)$ symmetry of the
elmentary four-dimensional solution of Dabholkar {\it et al.} \dghrr.
In their solution $T_1$ and $T_2$ are zero, but $\eta$ and $B_{\mu\nu}$
are non-zero. The relevant part of the action is
\eqn\dabactt{S_4={1\over 2\kappa_4^2}\int d^4 x\sqrt{-g}
\left( R - 2g^{\mu\nu}\partial_\mu \eta \partial_\nu \eta
-{1\over 12} e^{-4\eta} H_{\mu\nu\rho} H^{\mu\nu\rho}
\right).}
The equations of motion of this theory also display an $SL(2,R)$ symmetry,
but this becomes manifest only after dualizing and introducing the axion
field $a$ via
\eqn\axfield{\sqrt{-g}g^{\mu\nu}\partial_\nu a=
{1\over 3!}\epsilon^{\mu\nu\rho\sigma} H_{\nu\rho\sigma} e^{-4\eta}.}
Then in terms of the complex field
\eqn\comps{\eqalign{S&=S_1 + iS_2 \cr
&=a + ie^{-2\eta} \cr}}
the Dabholkar {\it et al.} fundamental string solution may be written
\eqn\dabs{\eqalign{S&={1\over 2\pi i} \ln {z\over r_0},\cr
ds^2&=-dt^2 + dx_5^2 - {1\over 2\pi} \ln {r\over r_0} dz d\overline z.\cr}}
Thus \tsol\ and \dabs\ are the same with the replacement $T\leftrightarrow
S$. It has been conjectured that this second $SL(2,Z)$ symmetry may also be a
symmetry of string theory \refs{\filq,\senone,\sch}, but this is far from
obvious order by order in the string loop expansion since it involves a
strong/weak coupling duality $\eta\to - \eta$. What interpretation
are we to give to these two $SL(2,Z)$ symmetries: one an obvious symmetry of
the fundamental string and the other an obscure symmetry of the fundamental
string?

While the present work was in progress, we became aware of recent
interesting papers by Sen \sentwo, Schwarz and Sen \schsen\ and Bin\'etruy
\bin. In particular, Sen draws attention to the Dabholkar {\it et al.} string
solution \dabs\ and its associated $SL(2,Z)$ symmetry as supporting evidence
in favor of the conjecture that $SL(2,Z)$ invariance may indeed be an exact
symmetry of string theory. He also notes
that the spectrum of electric and magnetic charges is consistent with the
proposed $SL(2,Z)$ symmetry \sentwo.\footnote{$^\dagger$}
{Sen also discusses the concept of a ``dual string", but for him this is
obtained from the fundamental string by an $SL(2,Z)$ transform. For us, a
dual string is obtained by the replacement $S\leftrightarrow T$.}

All of these observations fall into place if one accepts the proposal of
Schwarz and Sen \schsen: {\it under string/fivebrane duality the roles of
the target-space duality and the strong/weak coupling duality are
interchanged !} This proposal is entirely consistent with an earlier one that
under string/fivebrane duality the roles of the $\sigma$-model loop
expansion and the string loop expansion are interchanged \cqdual. In this
light, the two $SL(2,Z)$ symmetries discussed above are just what one
expects. From the string point of view, the $T$-field $SL(2,Z)$ is an
obvious target space symmetry, manifest order by order in string loops
whereas the $S$-field $SL(2,Z)$ is an obscure strong/weak coupling symmetry.
{}From the fivebrane point of view, it is the $T$-field $SL(2,Z)$ which is
obscure while the $S$-field $SL(2,Z)$ is an ``obvious" target space
symmetry.
(This has not yet been proved except at the level of the low-energy field
theory, however. It would be interesting to have a proof starting from the
worldvolume of the fivebrane.)
This interchange in the roles of the $S$ and $T$ field in
going from the string to the fivebrane has also been noted by Bin\'etruy
\bin. It is made more explicit when $S$ is expressed in terms of the
variables appearing naturally in the fivebrane version
\eqn\fbvers{\eqalign{S&=S_1 + iS_2 \cr
&=a_{056789} + ie^{-2\eta},\cr
&=a_{056789} + i\sqrt{{\rm det} g^F_{mn}}, \qquad m,n=3,4,6,7,8,9, \cr}}
where $g^F_{MN}=e^{-\phi/3}g^S_{MN}$ is the fivebrane $\sigma$-model metric
\duality\ and
$a_{MNPQRS}$ is the 6-form which couples to the 6-dimensional worldvolume of
the fivebrane, in complete analogy with \tfield.

Note, however, that unlike the Dabholkar {\it et al.} solution, our
symmetric solution \stsol\ also involves the non-abelian gauge fields $A_\rho,
\Phi,\Psi$ whose interactions appear to destroy the $SL(2,Z)$. This remains
a puzzle. (A generalization of the $D=4$ Dabholkar {\it et al.} solution
involving gauge fields may also be possible by obtaining it as a soliton of
the fivebrane theory. This would involve a $D=4$ analogue of the $D=10$
solution discussed in \dfluhs.)

\newsec{Discussion}

It may at first sight seem strange that a string can be dual to another
string in $D=4$. After all, the usual formula relating the dimension of an
extended object, $d$, to that of the dual object, $\tilde d$, is
$\tilde d=D-d-2$. So one might expect string/string duality only in $D=6$
\cqdual. However, when we compactify $n$ dimensions and allow
the dual object to wrap around $m\leq d-1$ of the compactified directions
we find $\tilde d_{{\rm effective}}=\tilde d -m=D_{{\rm effective}}-d-2
+(n-m)$, where $D_{{\rm effective}}=D-n$. In particular for $D_{{\rm
effective}}=4$, $d=2$, $n=6$ and $m=4$, we find $\tilde d_{{\rm
effective}}=2$.

Thus the whole string/fivebrane duality conjecture is put in a different
light when viewed from four dimensions. After all, our understanding of the
quantum theory of fivebranes in $D=10$ is rather poor, whereas the quantum
theory of strings in $D=4$ is comparatively well-understood (although we
still have to worry about the monopoles and domain walls). In particular,
the dual string will presumably exhibit the normal kind of mass spectrum
with linearly rising Regge trajectories, since the classical
($\hbar$-independent) string expression $\widetilde T_6 L^4 \times
({\rm angular\ momentum})$ has dimensions of $({\rm mass})^2$, whereas
the analogous classical expression for an uncompactified fivebrane is
$(\widetilde T_6)^{1/5} \times ({\rm angular\ momentum})$ which has
dimensions $({\rm mass})^{6/5}$ \duff. Indeed, together with the observation
that the $SL(2,Z)$ strong/weak coupling duality appears only after
compactifying at least $6$ dimensions, it is tempting to revive the earlier
conjecture \refs{\duff,\fujku}\ that the internal consistency of the
fivebrane may actually {\it require} compactification.

\bigbreak\bigskip\bigskip\centerline{{\bf Acknowledgements}}\nobreak
We would like to thank Pierre Bin\'etruy, Ruben Minasian, Joachim Rahmfeld,
John Schwarz and Ashoke Sen for helpful discussions.

\vfil\eject
\listrefs
\bye